\documentclass[aps,pra,preprint,onecolumn,superscriptaddress]{revtex4-1}
\usepackage{amsfonts}
\usepackage{amsmath}
\usepackage{amssymb}
\usepackage[colorlinks, citecolor=blue,linkcolor=red]{hyperref}
\usepackage{color}
\usepackage{float}
\usepackage{hyperref}
\usepackage{textcomp}
\usepackage{graphicx}

\begin{document}

\title{Symmetry breaking of spatial Kerr solitons in fractional dimension}
\author{Pengfei Li}
\email{lpf281888@gmail.com}
\address{Department of Physics, Taiyuan Normal University, Jinzhong, 030619, China}
\address{Institute of Computational and Applied Physics, Taiyuan Normal University, Jinzhong, 030619, China}
\author{Boris A. Malomed}
\address{Department of Physical Electronics, School of Electrical Engineering, Faculty of Engineering, and Center for Light-Matter Interaction, Tel Aviv University, Tel Aviv 69978, Israel}
\author{Dumitru Mihalache}
\address{Horia Hulubei National Institute of Physics and Nuclear Engineering, Magurele, Bucharest, RO-077125, Romania}

\begin{abstract}
We study symmetry breaking of solitons in the framework of a nonlinear
fractional Schr\"{o}dinger equation (NLFSE), characterized by its L\'{e}vy
index, with cubic nonlinearity and a symmetric double-well potential.
Asymmetric, symmetric, and antisymmetric soliton solutions are found, with
stable asymmetric soliton solutions emerging from unstable symmetric and
antisymmetric ones by way of symmetry-breaking bifurcations. Two different
bifurcation scenarios are possible. First, symmetric soliton solutions
undergo a symmetry-breaking bifurcation of the pitchfork type, which gives
rise to a branch of asymmetric solitons, under the action of the
self-focusing nonlinearity. Second, a family of asymmetric solutions
branches off from antisymmetric states in the case of self-defocusing nonlinearity
through a bifurcation of an inverted-pitchfork type. Systematic numerical
analysis demonstrates that increase of the L\'{e}vy index leads to shrinkage
or expansion of the symmetry-breaking region, depending on parameters of the
double-well potential. Stability of the soliton solutions is explored
following the variation of the L\'{e}vy index, and the results are confirmed
by direct numerical simulations.
\end{abstract}

\maketitle

\section{Introduction}

Spontaneous symmetry breaking (SSB), i.e., self-organized transformation of
symmetric or antisymmetric states into asymmetric ones, is a ubiquitous
phenomenon that occurs in a wide variety of intrinsically symmetric physical
systems. Examples of the SSB have been investigated theoretically in
nonlinear optics \cite{SSB-Review-Malomed}, Bose-Einstein condensates (BECs)
\cite{SSB1,SSB2,SSB3,SSB4}, lasers \cite{SSB5}, liquid crystals \cite{SSB6},
and other areas. In optics, the SSB has been observed, in particular, in a CS%
$_{2}$ planar optical waveguide \cite{SSBe1}, in a biased photorefractive
crystal (SBN:60) illuminated by a probe beam modulated by an amplitude mask
\cite{SSBe2}, in symmetrically coupled lasers \cite{SSBe3,NatPhot2},
metamaterials \cite{Kivshar}, etc. In BEC, the SSB was reported in the form
of macroscopic quantum self-trapping with an imbalanced population, in a
condensate of $^{87}$Rb atoms loaded in a symmetric double-well potential
\cite{SSBe4}. Especially, as one of fundamental aspects of the
optical-soliton phenomenology, the SSB of solitons was intensively
investigated in various settings modeled by nonlinear Schr\"{o}dinger
equations (NLSEs) \cite%
{SSB-OptSol1,SSB-OptSol2,SSB-OptSol3,SSB-OptSol4,SSB-OptSol5,SSB-OptSol6,
SSB-OptSol7,SSB-OptSol8,SSB-OptSol9,SSB-OptSol10,SSB-OptSol11,SSB-OptSol12,
SSB-OptSol13,SSB-OptSol14,SSB-OptSol15,SSB-OptSol16,NatPhot1,Elad}.

Recently, the investigation of SSB of solitons has been expanded into
non-Hermitian optical systems, which are modeled by NLSEs with parity-time ($%
\mathcal{PT}$) symmetric complex potentials \cite{PT1,PT2}, where families
of stable asymmetric one-dimensional (1D) solitons induced by the SSB have
been found for specially designed complex potentials \cite%
{YJK1,YJK2,nonPT-Li,YJK3}, as well as under the action of 2D potentials \cite%
{nonPT-2D}.

An interesting generalization of the Schr\"{o}dinger equation, corresponding
to fractional-dimensional Hamiltonians, was proposed in Refs. \cite{FSE2}-%
\cite{FSE1}. It has been introduced, in the context of the quantum theory,
via Feynman path integrals over L\'{e}vy-flight trajectories, which leads to
the fractional Schr\"{o}dinger equation (FSE). Although implications of such
models are still a matter of debate \cite{FSE4,FSE5}, some experimental
schemes have been proposed for emulating them in condensed-matter settings
and optical cavities \cite{FSE11,FSE12}. A number of intriguing dynamical
properties have been predicted in the framework of FSEs, including 1D zigzag
propagation \cite{FSE13}, diffraction-free beams \cite%
{FSE14,FSE15,FSE16,FSE17,FSE18}, beam splitting \cite{FSE19}, periodic
oscillations of Gaussian beams \cite{FSE20}, beam-propagation management
\cite{FSE21}, optical Bloch oscillations and Zener tunneling \cite{FSE22},
resonant mode conversion and Rabi oscillations \cite{FSE23}, localization
and Anderson delocalization \cite{FSE24}, and SSB of $\mathcal{PT}$%
-symmetric modes in a linear FSE \cite{FSE25}.

Naturally, FSEs may be applied to nonlinear-optical settings \cite%
{NLFSE2,NLFSE3}. Recent works demonstrate that a variety of fractional
optical solitons can be produced by nonlinear FSEs (NLFSEs) \cite{NLFSE4},
such as ``accessible solitons" \cite{NLFSE5,NLFSE6,NLFSE7}, double-hump
solitons and fundamental solitons in $\mathcal{PT}$-symmetric potentials
\cite{NLFSE8,NLFSE9}, gap solitons and surface gap solitons in $\mathcal{PT}$%
-symmetric photonic lattices \cite{NLFSE10,NLFSE11}, two-dimensional
solitons \cite{NLFSE12}, and off-site- and on-site-centered vortex solitons
in two-dimensional $\mathcal{PT}$-symmetric lattices \cite{NLFSE13}.
Composition relations between nonlinear Bloch waves and gap solitons have
also been addressed in the framework of NLFSE with photonic lattices \cite%
{NLFSE14}, bright solitons under the action of periodically spatially
modulated Kerr nonlinearity \cite{NLFSE15}, as well as dissipative surface
\cite{NLFSE16} and bulk solitons \cite{Yingji}.

Once solitons may undergo various forms of SSB in models based on
conventional Hermitian and $\mathcal{PT}$-symmetric NLSEs, it is natural to
consider the possibility of similar effects to occur in NLFSEs with
symmetric double-well potentials. In this work, we address SSB of solitons
in NLFSEs with self-focusing and self-defocusing cubic nonlinearities. The
model and numerical methods are introduced in Sec. II. Basic numerical
results, including the existence of soliton solutions in the fractional
dimension, their SSB bifurcations, and results for stability and dynamics of
the solitons are reported in Sec. III. The paper is concluded by Sec. IV.

\section{The model and methods}

\subsection{The nonlinear fractional Schr\"{o}dinger equation}

We start the analysis by considering beam propagation along the $z$-axis in
a waveguide with the Kerr nonlinearity. The balance between the diffraction
in the fractional dimension \cite{FSE12}, cubic nonlinearity, and modulation
of the linear refractive index makes it possible to form solitons. The
corresponding model for the light-beam propagation is based on the following
equation:%
\begin{equation}
i\frac{\partial A}{\partial z}-\frac{1}{2kw_{0}^{2}}\left( -\frac{\partial
^{2}}{\partial \xi ^{2}}\right) ^{\alpha /2}A+\frac{k\left[ n\left( x\right)
-n_{0}\right] }{n_{0}}A+k_{0}n_{2}\left\vert A\right\vert ^{2}A=0,  \label{NLFSE1}
\end{equation}%
where $A(z,x)$ is the local amplitude of the optical field, and $\xi =x/w_{0}
$ is the normalized transverse coordinate, scaled to characteristic width $%
w_{0}$ of the input beam. The fractional derivative, $(-\partial
^{2}/\partial \xi ^{2})^{\alpha /2}$, is determined by the L\'{e}vy index,
which usually takes values $1\leq \alpha \leq 2$ (this point is additionally
considered below). Further, $k=k_{0}n_{0}$ is\ the wavenumber, with $n_{0}$
being the background refractive index, and $k_{0}=2\pi /\lambda $, where $%
\lambda $ is the optical wavelength. An effective potential is introduced by
local modulation, $n(x)$, of the linear refractive index, while $n_{2}$ is
the nonlinear Kerr coefficient of the material.

Equation (\ref{NLFSE1}) can be cast in a normalized form by means of
additional rescaling, $\Psi (\zeta ,\xi )\equiv \left( k_{0}\left\vert
n_{2}\right\vert L_{d}\right) ^{1/2}A(z,x)$, where $L_{d}=kw_{0}^{2}$ is the
diffraction length, $\zeta =z/L_{d}$ is the normalized coordinate in the
propagating direction, and the effective potential is defined as $-V(\xi
)\equiv -L_{d}k\left[ n\left( w_{0}\xi \right) -n_{0}\right] /n_{0}$. The
scaled equation is%
\begin{equation}
i\frac{\partial \Psi }{\partial \zeta }-\frac{1}{2}\left( -\frac{\partial
^{2}}{\partial \xi ^{2}}\right) ^{\alpha /2}\Psi +V\left( \xi \right) \Psi
+\sigma \left\vert \Psi \right\vert ^{2}\Psi =0,  \label{NLFSE2}
\end{equation}%
where $\sigma \equiv n_{2}/\left\vert n_{2}\right\vert =\pm 1$ corresponds
to the self-focusing $(+)$ and defocusing $(-)$ nonlinearity, respectively.
Obviously, when $\alpha =2$, Eq. (\ref{NLFSE2}) amounts to the standard
NLSE, and when $\alpha =1$, the fractional derivative corresponds to the
operator in the form of the ``square root of Laplacian\textquotedblright ,
which was introduced long ago in a phenomenological model of instability of
combustion fronts \cite{Aldushin}.

In this paper, we study SSB for soliton solutions to Eq. (\ref{NLFSE2}),
chiefly with $1\leq \alpha <2$, although the case of $\alpha <1$ is briefly
considered too, see Fig. \ref{figure3} below. To this end, we introduce the
effective symmetric double-well potential corresponding to%
\begin{equation}
V\left( \xi \right) =V_{0}\left\{ \exp \left[ -\left( \frac{\xi +\xi _{0}}{%
W_{0}}\right) ^{2}\right] +\exp \left[ -\left( \frac{\xi -\xi _{0}}{W_{0}}%
\right) ^{2}\right] \right\} ,  \label{Potential}
\end{equation}%
with potential minima set at points $\xi =\pm \xi _{0}$, while $W_{0}$ and $%
V_{0}$ denote the width and depth of the local wells.

Soliton solutions produced by Eq. (\ref{NLFSE2}) with real propagation
constant $\beta $ are sought as%
\begin{equation}
\Psi \left( \zeta ,\xi \right) =\psi \left( \xi \right) e^{i\beta \zeta },
\label{Solu}
\end{equation}%
where the real function $\psi (\xi )$ obeys a stationary equation
\begin{equation}
-\frac{1}{2}\left( -\frac{d^{2}}{d\xi ^{2}}\right) ^{\alpha /2}\psi +V\left(
\xi \right) \psi +\sigma \left\vert \psi \right\vert ^{2}\psi -\beta \psi =0.
\label{NLFSE3}
\end{equation}%
Soliton solutions of Eq. (\ref{NLFSE3}) are characterized by the integral
power (norm), which is a dynamical invariant of Eq. (\ref{NLFSE2}), and may
be naturally split in contributions from the left and right regions:
\begin{equation}
P(\beta )\equiv P_{L}(\beta )+P_{R}(\beta )=\int\nolimits_{-\infty
}^{+\infty }\left\vert \Psi \right\vert ^{2}d\xi \equiv
\int\nolimits_{-\infty }^{0}\left\vert \Psi \right\vert ^{2}d\xi
+\int\nolimits_{0}^{+\infty }\left\vert \Psi \right\vert ^{2}d\xi .
\label{P}
\end{equation}%
Asymmetric states are then characterized by the SSB parameter
\begin{equation}
\Theta =\frac{P_{L}-P_{R}}{P_{L}+P_{R}},  \label{Theta}
\end{equation}%
which is used below.

\subsection{The numerical method}

The fractional derivative in Eq. (\ref{NLFSE3}) is defined as a
pseudo-differential operator \cite{JMP51-062102,JMP54-012111}
\begin{equation}
\mathcal{F}[\left( -d^{2}/d\xi ^{2}\right) ^{\alpha /2}\psi ]=\left\vert
k\right\vert ^{\alpha }\mathcal{F}\psi (k),  \label{Fourier}
\end{equation}%
where $\mathcal{F}$ denotes the Fourier transform, which converts functions
of $x$ into functions of the respective wavenumber, $k$. Actually, for all
values $\alpha <2$, Eq. (\ref{NLFSE3}) is a nonlocal equation. Several
methods have been developed for handling nonlocal FSEs \cite%
{JCP325,CMA71,CMA75}.

To obtain numerical soliton solutions of Eq. (\ref{NLFSE3}), we employed
the\ Newton-conjugate-gradient method, as presented in Refs. \cite{NCG,Book1}%
. Accordingly, Eq. (\ref{NLFSE3}) is rewritten as%
\begin{equation}
L_{0}\psi \left( \xi \right) =0,  \label{NCG1}
\end{equation}%
where%
\begin{equation}
L_{0}=-\frac{1}{2}\left( -\frac{d^{2}}{d\xi ^{2}}\right) ^{\alpha /2}+V(\xi
)+\sigma \left\vert \psi \right\vert ^{2}-\beta .  \label{NCG2}
\end{equation}%
Here, the propagation constant $\beta $ is considered as a parameter with a
fixed value, while solution $\psi (\xi )$ is generated by means of Newton's
iterations
\begin{equation}
\psi _{n+1}=\psi _{n}+\Delta \psi _{n},  \label{NCG3}
\end{equation}%
where $\psi _{n}$ is an approximate solution, and correction $\Delta \psi
_{n}$ is computed from the linear Newton's equation
\begin{equation}
L_{1n}\Delta \psi _{n}=-L_{0}\left( \psi _{n}\right) ,  \label{NCG4}
\end{equation}%
where $L_{1n}$ is the linearization operator $L_{1}$ corresponding to Eq. (%
\ref{NCG1}), evaluated with $\psi $ replaced by approximate solution $\psi
_{n}$:%
\begin{equation}
L_{1}=-\frac{1}{2}\left( -\frac{d^{2}}{d\xi ^{2}}\right) ^{\alpha
/2}+V+\sigma \left\vert \psi \right\vert ^{2}+2\sigma \psi \mathtt{Re}\left(
\psi ^{\ast }\right) -\beta .  \label{NCG5}
\end{equation}%
In the framework of this scheme, Eq. (\ref{NCG4}) can be solved directly by
dint of conjugate gradient iterations \cite{NCG,Book1}, which yields
symmetric, antisymmetric, and asymmetric soliton solutions.

\section{Numerical results}

\subsection{Symmetric, antisymmetric, and asymmetric soliton solutions in
the fractional dimension}

\begin{figure}[tbp]
\centering\vspace{0cm} \includegraphics[width=8.5cm]{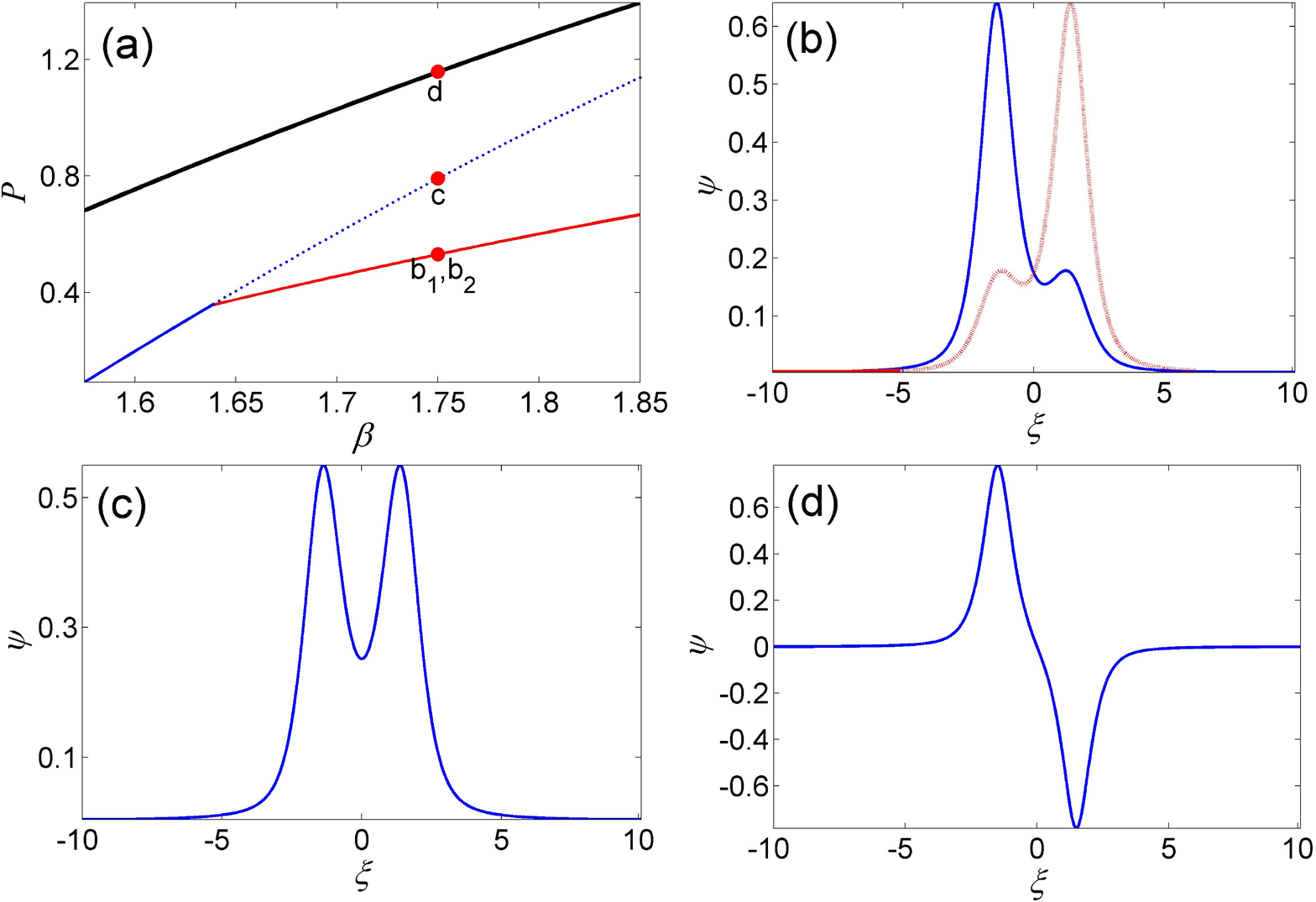} \vspace{%
0.0cm}
\caption{(Color online) SSB of solitons of Eq. (\protect\ref{NLFSE2}) with L%
\'{e}vy index $\protect\alpha =1.1$, self-focusing Kerr nonlinearity ($%
\protect\sigma =+1$), and the double-well potential (\protect\ref{Potential}%
) with parameters $V_{0}=2$, $\protect\xi _{0}=1.5$, and $W_{0}=1.4$.\ (a)
Power curves of families of stable and unstable symmetric solitons (thin
solid and dotted blue lines, respectively), stable antisymmetric solitons
(the thick solid black line), and stable asymmetric solitons (the thin solid
red line). Panels (b), (c), and (d) display asymmetric, symmetric, and
antisymmetric soliton solutions at $\protect\beta =1.75$, marked by ``b$_{1}$%
,b$_{2}$,c,d\textquotedblright\ in panel (a).}
\label{figure1}
\end{figure}

\begin{figure}[tbp]
\centering\vspace{0cm} \includegraphics[width=8.5cm]{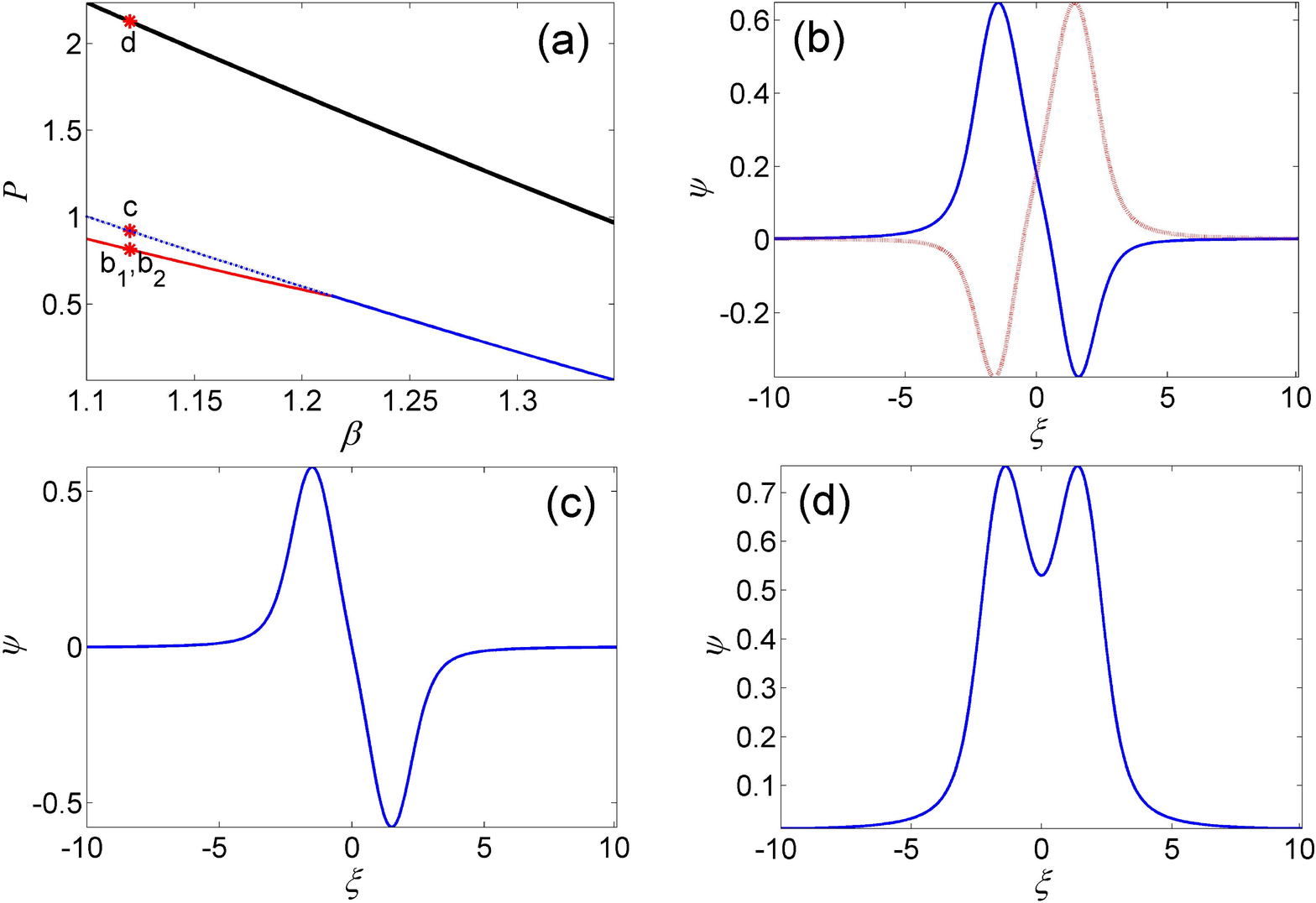} \vspace{%
0.0cm}
\caption{(Color online) SSB of solitons of Eq. (\protect\ref{NLFSE2}) with
self-defocusing Kerr nonlinearity ($\protect\sigma =-1$), other parameters
being the same as in Fig. \protect\ref{figure1}.\ (a) Power curves of stable
symmetric solitons (the thick solid black line), unstable and stable
antisymmetric solitons (thin dotted and solid blue lines, respectively), and
stable asymmetric solitons (the thin solid red line). Panels (b), (c), and
(d) depict asymmetric, antisymmetric, and symmetric solitons at $\protect%
\beta =1.12$, marked by ``b$_{1}$,b$_{2}$,c,d\textquotedblright\ in panel
(a).}
\label{figure2}
\end{figure}

Generic examples of soliton families are presented in Fig. \ref{figure1}.
They were found as numerical solutions of Eq. (\ref{NLFSE3}) with L\'{e}vy
index $\alpha =1.1$ and $\sigma =+1$ (the self-focusing sign of the
nonlinearity), while the potential $V\left( \xi \right) $ is taken as per
Eq.\ (\ref{Potential}) with $V_{0}=2$, $\xi _{0}=1.5$, and $W_{0}=1.4$.
Figure \ref{figure1}(a) shows power curves $P(\beta )$ for stable and
unstable symmetric-soliton solutions, as well as for stable antisymmetric
and stable asymmetric ones in this potential. SSB occurs when the integral
power of symmetric solitons exceeds a critical value (which corresponds to
the bifurcation point), $P_{\mathrm{cr}}\approx 0.3603$, the respective
propagation constant being $\beta _{\mathrm{cr}}\approx 1.6389$. As a result
of the SSB, the power curves of the symmetric and asymmetric soliton
solutions form a pitchfork bifurcation. As examples, profiles of asymmetric,
symmetric, and antisymmetric soliton solutions with $\beta =1.75$ are
displayed in Figs. \ref{figure1}(b), \ref{figure1}(c), and \ref{figure1}(d),
respectively. Panel \ref{figure1}(b) demonstrates that there actually exist
two branches of the asymmetric soliton solutions, which are mirror images of
each other, $\psi \left( \xi \right) $ and $\psi \left( -\xi \right) $,
represented by coinciding points b$_{1}$ and b$_{2}$\ in Fig. \ref{figure1}%
(a).

SSB of solitons is also found in Eq. (\ref{NLFSE3}) with the self-defocusing
nonlinearity ($n_{2}<0$), when the symmetric soliton solutions, shown in
Fig. \ref{figure2}(a), are stable, but the antisymmetric ones become
unstable at $P_{\mathrm{cr}}\approx 0.5459$ ($\beta _{\mathrm{cr}}\approx
1.2138$) with the increase of the integral power, as is also shown in Fig. %
\ref{figure2}(a), where a branch of stable asymmetric soliton solutions
bifurcates from the\ branch of the unstable antisymmetric solutions. Figures %
\ref{figure2}(b), \ref{figure2}(c), and \ref{figure2}(d) display the
corresponding profiles of the asymmetric, symmetric, and antisymmetric
soliton solutions, respectively, with $\beta =1.12$. In this case too, there
are also two branches of asymmetric states, which are mirror images of each
other, $\psi (\xi )$ and $\psi (-\xi )$, as can be seen in panel \ref%
{figure2}(b).

\begin{figure}[tbp]
\centering\vspace{0cm} \includegraphics[width=8.5cm]{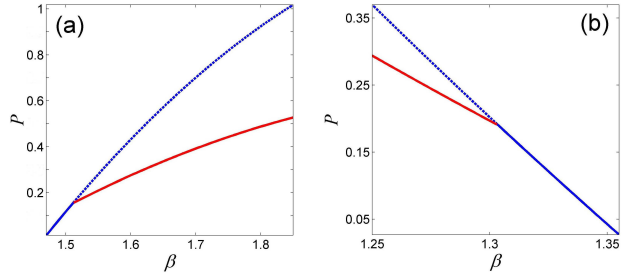} \vspace{%
0.0cm}
\caption{(Color online) SSB of solitons of Eq. (\protect\ref{NLFSE2}) with L%
\'{e}vy index $\protect\alpha =0.8$, other parameters being the same as in
Fig. \protect\ref{figure1}.\ (a) Power curves of families of stable and
unstable symmetric solitons (solid and dotted blue lines, respectively), and
stable asymmetric solitons (the solid red line) with the self-focusing Kerr
nonlinearity. (b) Power curves of unstable and stable antisymmetric solitons
(dotted and solid blue lines, respectively), and stable asymmetric solitons
(the solid red line) with the self-defocusing Kerr nonlinearity.}
\label{figure3}
\end{figure}

In fact, even in the case of $\alpha <1$, when solitons, supported by the
self-focusing cubic nonlinearity in the free space, are unstable, because of
the occurrence of the supercritical collapse \cite{Yingji}, the trapping
potential may stabilize the solitons and uphold the familiar SSB scenario,
as shown in Fig. \ref{figure3}(a) for $\alpha =0.8$. In the case of the
self-defocusing nonlinearity, values $\alpha <1$ are possible too, giving
rise to the same type of SSB as considered above, see Fig. \ref{figure3}(b).

The fact that the growth of the strength of the focusing or defocusing
nonlinearity destabilizes, respectively, symmetric or antisymmetric
solitons, replacing them by asymmetric ones, is a generic property of SSB
\cite{SSB-Review-Malomed}. It is relevant to mention that, in the latter
case, the pitchfork bifurcation may be identified as an \textit{inverted}
one, because it takes place, in Figs. \ref{figure2}(a) and \ref{figure3}(b),
with the \textit{decrease} of the propagation constant (but still with the
increase of the integral power). It is also found that, in the case of the
self-focusing nonlinearity, the stable soliton branches in Figs. \ref%
{figure1}(a) and \ref{figure3}(a) satisfy the Vakhitov-Kolokolov (VK)
criterion \cite{VK1,VK2}, and, for the defocusing nonlinearity, the stable
soliton solution branches in Figs. \ref{figure2}(a) and \ref{figure3}(b)
satisfy the anti-VK criterion \cite{anti-VK}.

\subsection{Symmetry-breaking bifurcations at different values of the L\'{e}%
vy index}

\begin{figure}[tbp]
\centering\vspace{0cm} \includegraphics[width=8.5cm]{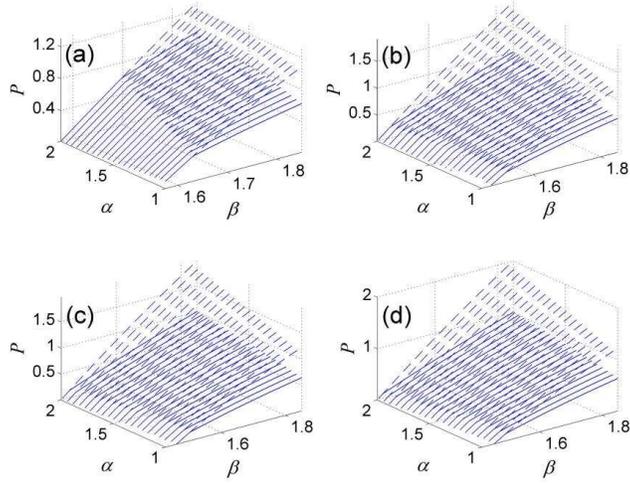} \vspace{0cm%
}
\caption{(Color online) The SSB bifurcation represented by dependences of
the integral power of unstable and stable symmetric (dashed and solid
curves, respectively) and stable asymmetric (solid curves) soliton solutions
on L\'{e}vy index $\protect\alpha $ and propagation constant $\protect\beta $
in the case of the self-focusing Kerr nonlinearity ($\protect\sigma =+1$)
for different values of the separation parameter of the double-well
potential (\protect\ref{Potential}): (a) $\protect\xi _{0}=1.5$, (b) $%
\protect\xi _{0}=1.8$, (c) $\protect\xi _{0}=1.9$, and (d) $\protect\xi %
_{0}=2$, respectively. Other parameters of the potential are the same as in
Fig. \protect\ref{figure1}.}
\label{figure4}
\end{figure}

\begin{figure}[tbp]
\centering\vspace{0cm} \includegraphics[width=8.5cm]{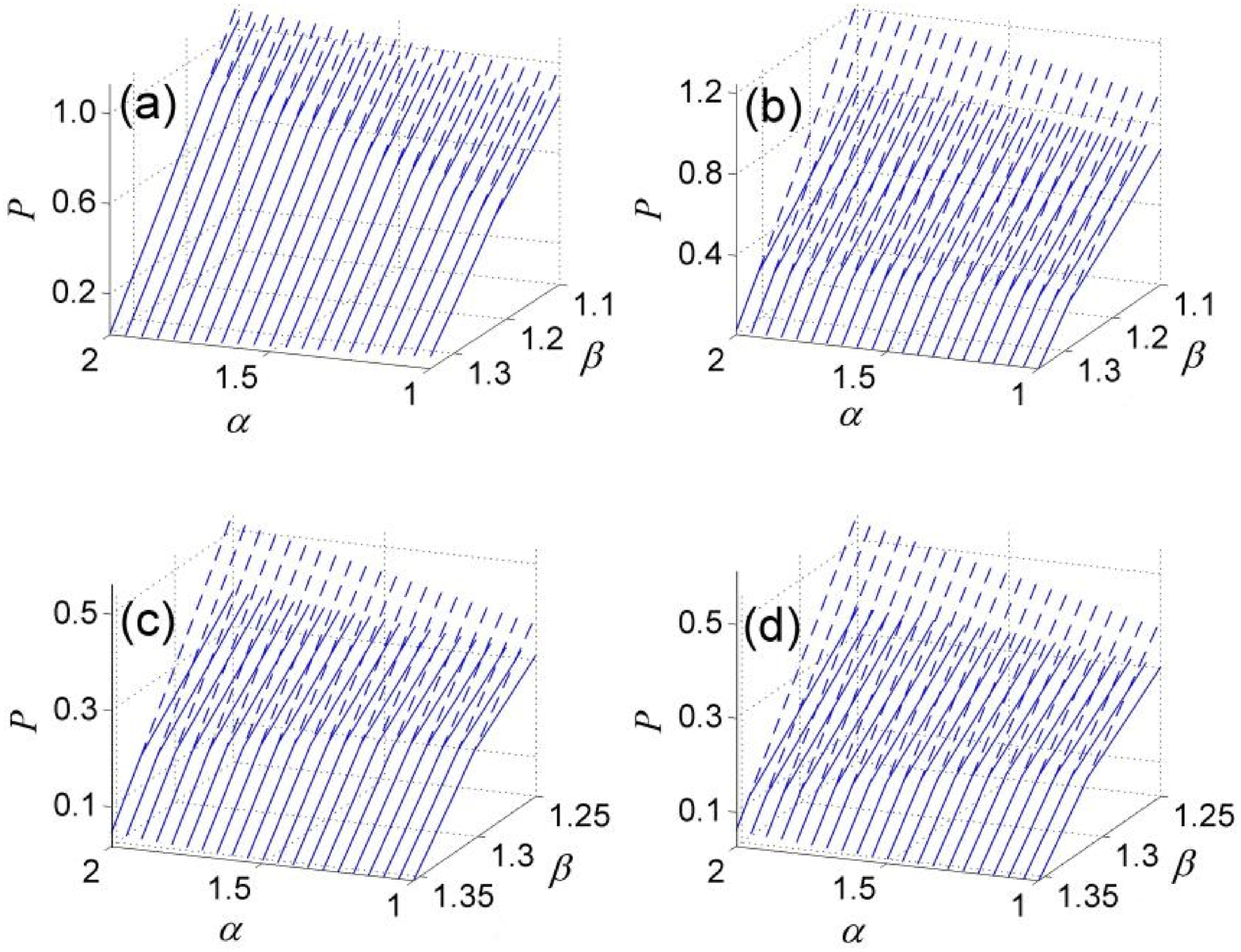} \vspace{0cm%
}
\caption{(Color online) The SSB bifurcation represented by dependences of
the integral power of unstable and stable antisymmetric (dashed and solid
curves, respectively) and stable asymmetric (solid curves) soliton solutions
on L\'{e}vy index $\protect\alpha $ and propagation constant $\protect\beta $
in the case of the self-defocusing Kerr nonlinearity ($\protect\sigma =-1$)
for different values of the separation parameter of the double-well
potential (\protect\ref{Potential}): (a) $\protect\xi _{0}=1.5$, (b) $%
\protect\xi _{0}=1.8$, (c) $\protect\xi _{0}=1.9$, and (d) $\protect\xi %
_{0}=2$, respectively. Other parameters of the potential are the same as in
Fig. \protect\ref{figure1}.}
\label{figure5}
\end{figure}

\begin{figure}[tbp]
\centering\vspace{0cm} \includegraphics[width=8.5cm]{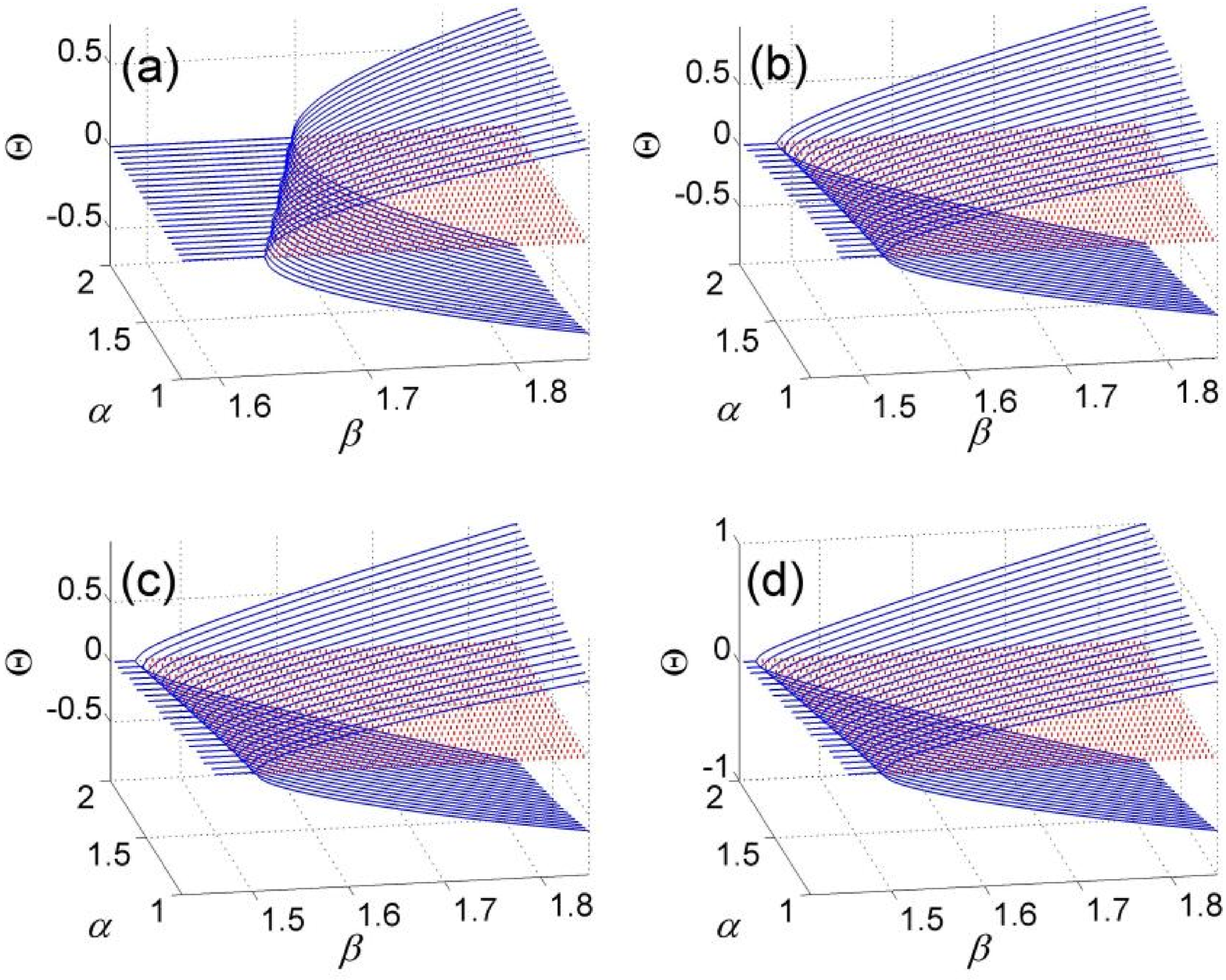} \vspace{0cm%
}
\caption{(Color online) SSB bifurcation diagrams shown in terms of asymmetry
parameter (\protect\ref{Theta}), corresponding to the situation displayed in
Fig. \protect\ref{figure4}: (a) $\protect\xi _{0}=1.5$, (b) $\protect\xi %
_{0}=1.8$, (c) $\protect\xi _{0}=1.9$, and (d) $\protect\xi _{0}=2$. Dashed
red curves represent unstable symmetric soliton solutions.}
\label{figure6}
\end{figure}

\begin{figure}[tbp]
\centering\vspace{0cm} \includegraphics[width=8.5cm]{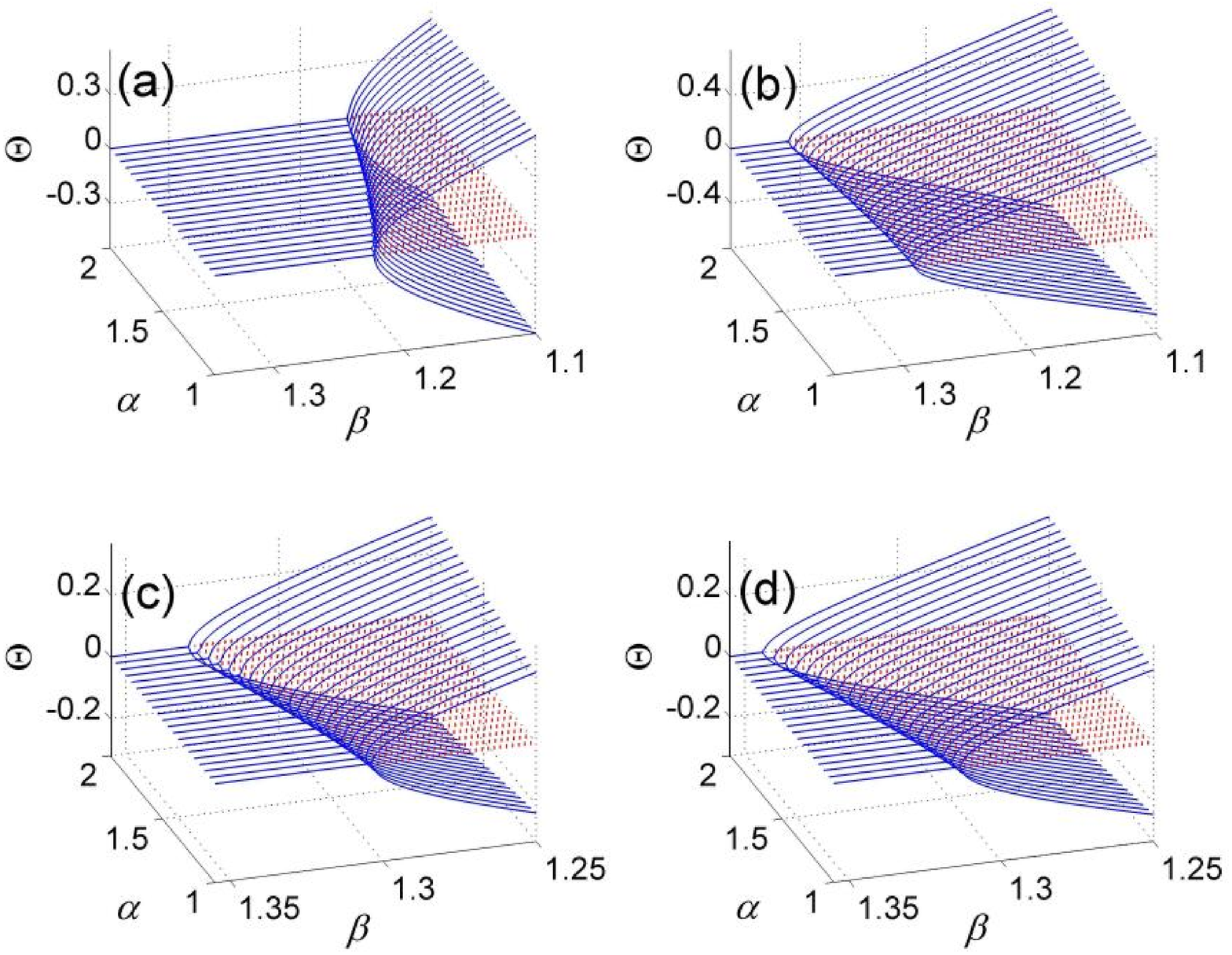} \vspace{0cm%
}
\caption{(Color online) SSB bifurcation diagrams shown in terms of asymmetry
parameter (\protect\ref{Theta}), corresponding to the situation displayed in
Fig. \protect\ref{figure5}: (a) $\protect\xi _{0}=1.5$, (b) $\protect\xi %
_{0}=1.8$, (c) $\protect\xi _{0}=1.9$, and (d) $\protect\xi _{0}=2$. Dashed
red curves represent unstable antisymmetric soliton solutions.}
\label{figure7}
\end{figure}

To explore how the value of L\'{e}vy index $\alpha $ effects the SSB
bifurcations, in Figs. \ref{figure4}, \ref{figure5}, \ref{figure6}, \ref%
{figure7}, \ref{figure8}, and \ref{figure9}, we have produced power curves
for soliton families, by numerically solving Eq. (\ref{NLFSE3}) and varying $%
\alpha $ (with interval of $0.05$, in the interval $1\leq \alpha \leq 2$)
and separation parameter $\xi _{0}$ of the double-well potential (\ref%
{Potential}). The numerical results for the self-focusing and defocusing
cases are summarized in Figs. \ref{figure4} and \ref{figure5}. Specifically,
at $\xi _{0}=1.5$, SSB areas in the plane of $\alpha $ and propagation
constant $\beta $ shrink with the increase of $\alpha $, as seen in Fig. \ref%
{figure4}(a) and \ref{figure5}(a). As $\xi _{0}$ increases to $1.8$, the
areas slightly expand with the increase of $\alpha $ in Fig. \ref{figure4}%
(b) and Fig. \ref{figure5}(b). Then, as $\xi _{0}$ increases to $1.9$ and $%
2.0$, the SSB\ areas significantly expand with the increase of $\alpha $ in
Figs. \ref{figure4}(c), \ref{figure4}(d), \ref{figure5}(c), and \ref{figure5}%
(d).

The families of the soliton solutions are further characterized by the
dependences of asymmetry parameter (\ref{Theta}) on $\alpha $ and $\beta $,
which are displayed in Figs. \ref{figure6} and Figs. \ref{figure7} for the
self-focusing and defocusing nonlinearities, respectively. The results
indicate that the SSB bifurcations in Eq. (\ref{NLFSE3}) are of the
supercritical (alias forward) type, i.e., with branches of asymmetric
branches going forward from the bifurcation point \cite{book}.

Further, Figs. \ref{figure6} and Figs. \ref{figure7} clearly demonstrate
that the increase of separation $\xi _{0}$ in the double-hump potential (\ref%
{Potential}) shifts the SSB bifurcation to lower values of the power. This
conclusion is quite natural, as a larger separation makes the effective
linear coupling between the two well weaker, making it easier for the
nonlinearity to compete with it \cite{SSB-Review-Malomed}.

\subsection{The linear stability analysis and dynamics}

\begin{figure}[tbp]
\centering\vspace{0cm} \includegraphics[width=8.5cm]{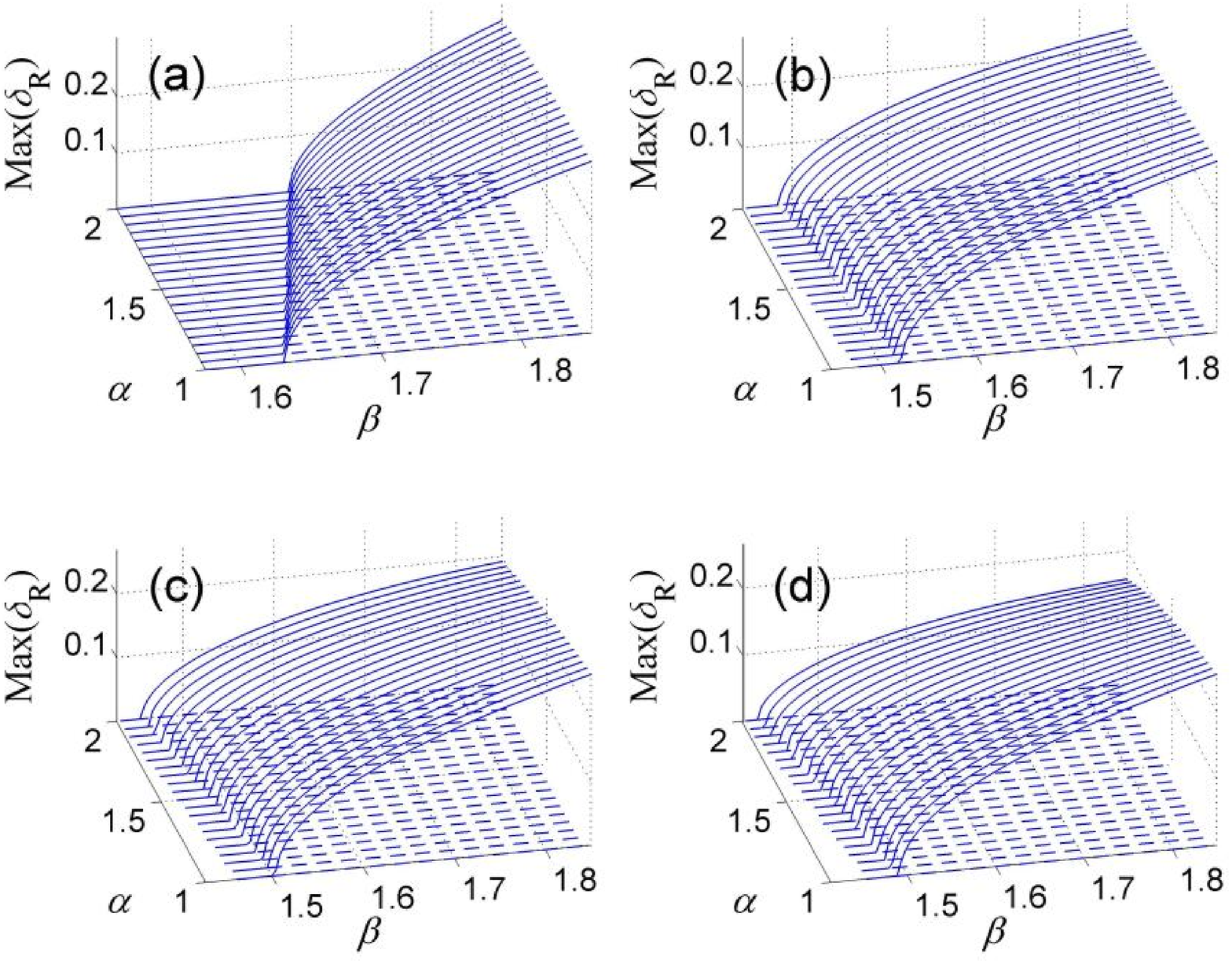} \vspace{0cm%
}
\caption{(Color online) Largest linear instability growth rates of symmetric
(solid) and asymmetric (dashed) soliton solutions from Fig. \protect\ref%
{figure4}.}
\label{figure8}
\end{figure}

\begin{figure}[tbp]
\centering\vspace{0cm} \includegraphics[width=8.5cm]{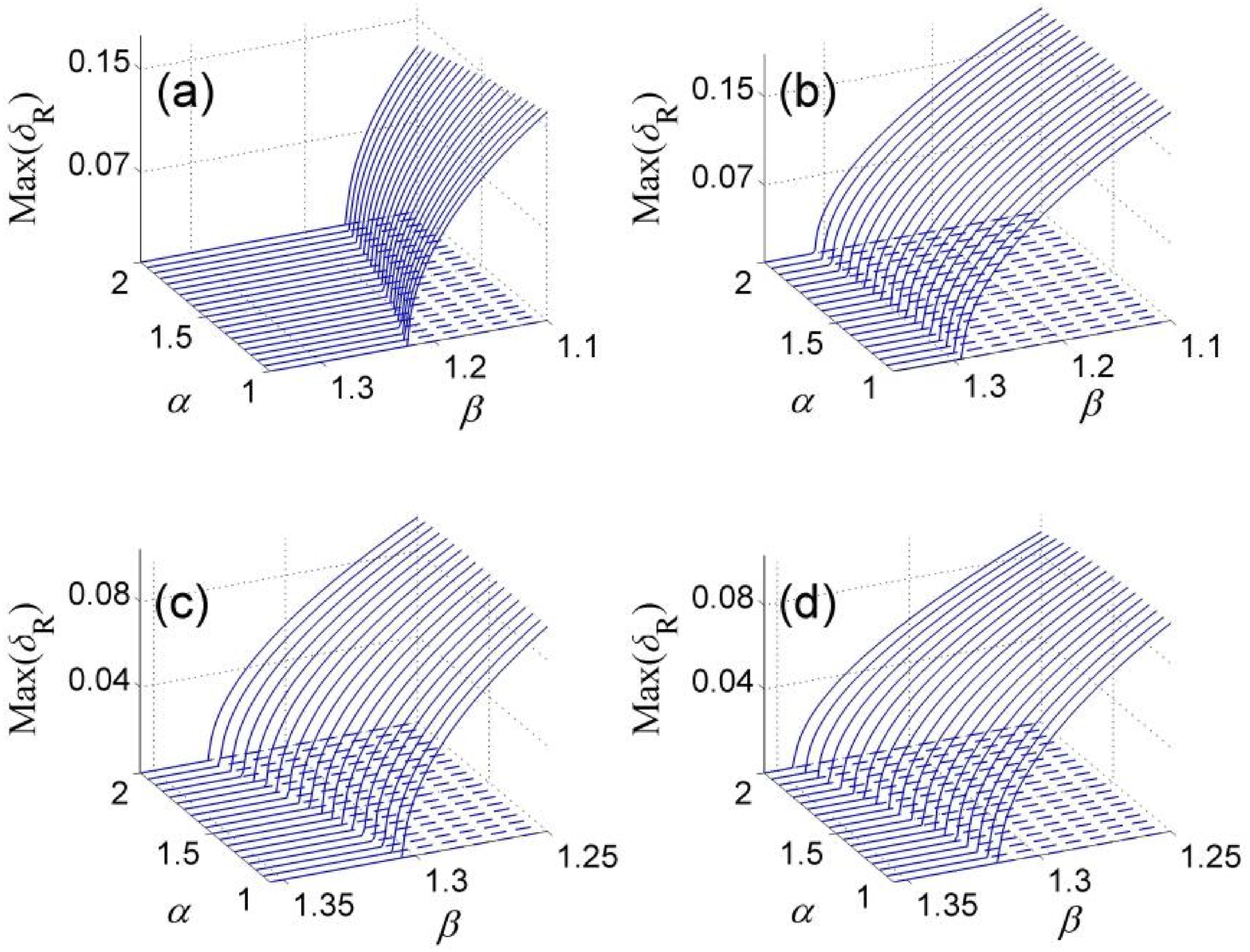} \vspace{0cm%
}
\caption{(Color online) Largest linear instability growth rates of
antisymmetric (solid) and asymmetric (dashed) soliton solutions in Fig.
\protect\ref{figure5}.}
\label{figure9}
\end{figure}

Stability of solitons in the present model was explored by considering small
perturbations $u\left( \xi \right) $ and $v\left( \xi \right) $ added to a
stationary solution, $\psi \left( \xi \right) $, with propagation constant $%
\beta $, as follows:%
\begin{equation}
\Psi \left( \zeta ,\xi \right) =e^{i\beta \zeta }\left[ \psi \left( \xi
\right) +u\left( \xi \right) e^{\delta \zeta }+v^{\ast }\left( \xi \right)
e^{\delta ^{\ast }\zeta }\right] ,  \label{Perturbation}
\end{equation}%
where $\delta $ is the instability growth rate (a complex one, in the
general case). Substituting this expression in Eq. (\ref{NLFSE2}) and
linearizing it with respect to the perturbations, we arrive at the following
linear eigenvalue problem:%
\begin{equation}
i\left(
\begin{array}{cc}
\mathcal{L}_{11} & \mathcal{L}_{12} \\
\mathcal{L}_{21} & \mathcal{L}_{22}%
\end{array}%
\right) \left(
\begin{array}{c}
u \\
v%
\end{array}%
\right) =\delta \left(
\begin{array}{c}
u \\
v%
\end{array}%
\right) ,  \label{LinearEigEqs}
\end{equation}%
where we define%
\begin{equation}
\mathcal{L}_{11}=-\frac{1}{2}\left( -\frac{d^{2}}{d\xi ^{2}}\right) ^{\alpha
/2}+U-\beta +2\sigma \left\vert \psi \right\vert ^{2},  \label{L11}
\end{equation}%
\begin{equation}
\mathcal{L}_{12}=+\sigma \psi ^{2},  \label{L12}
\end{equation}%
\begin{equation}
\mathcal{L}_{21}=-\sigma \psi ^{2},  \label{L21}
\end{equation}%
\begin{equation}
\mathcal{L}_{22}=\frac{1}{2}\left( -\frac{d^{2}}{d\xi ^{2}}\right) ^{\alpha
/2}-U+\beta -2\sigma \left\vert \psi \right\vert ^{2}.  \label{L22}
\end{equation}%
The underlying stationary solution is unstable if the solution of Eq. (\ref%
{LinearEigEqs}) produces at least one eigenvalue with $\mathrm{Re}(\delta
)>0 $.

Because Eq. (\ref{LinearEigEqs}) contains the fractional derivative, it is
convenient to solve it by means of the Fourier decomposition, which converts
the equation into a matrix eigenvalue problem for Fourier coefficients of
perturbation eigenfunctions $u\left( \xi \right) $ and $v\left( \xi \right) $%
. Then, the spectrum can be obtained numerically, using the Fourier
collocation and Newton-conjugate-gradient methods \cite{Book1,Book2,Book3}.
Both methods have produced identical results.

Figures \ref{figure8} and \ref{figure9} display the largest instability
growth rates, $\max \left( \delta _{R}\right) $, for symmetric and
asymmetric, or antisymmetric and asymmetric, sets of solitons, in the cases
of the self-focusing and defocusing nonlinearity, respectively. In agreement
with general principles \cite{book,SSB-Review-Malomed}, the SSB destabilizes
the symmetric or antisymmetric solitons in the former and latter cases,
respectively, while the asymmetric solitons remain stable solutions in the
entire interval of values of the L\'{e}vy index considered in the present
work, i.e., $1\leq \alpha \leq 2$.

\begin{figure}[tbp]
\centering\vspace{0cm} \includegraphics[width=8.5cm]{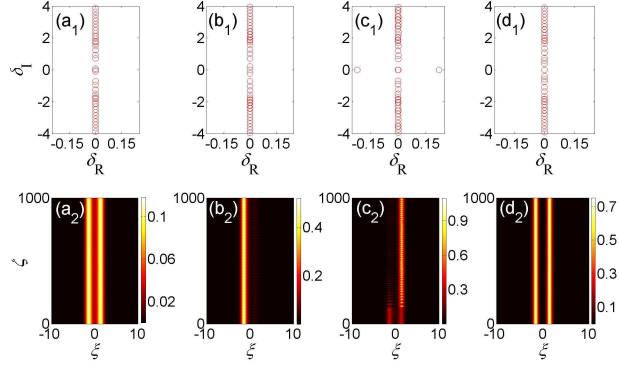} \vspace{%
0cm}
\caption{(Color online) The top row: linear-stability spectra for the
soliton solutions in the case of the self-focusing nonlinearity, $\protect%
\sigma =+1$. The bottom row: results of simulations of the perturbed
evolution of the solitons from the top row. (a$_{1}$,a$_{2}$): A stable
symmetric soliton with $\protect\beta =1.62$. Panels (b$_{1}$,b$_{2}$), (c$%
_{1}$,c$_{2}$), and (d$_{1}$,d$_{2}$) display the results for stable
asymmetric, unstable symmetric, and stable antisymmetric solitons,
respectively, at $\protect\beta =1.75$.}
\label{figure10}
\end{figure}

\begin{figure}[tbp]
\centering\vspace{0cm} \includegraphics[width=8.5cm]{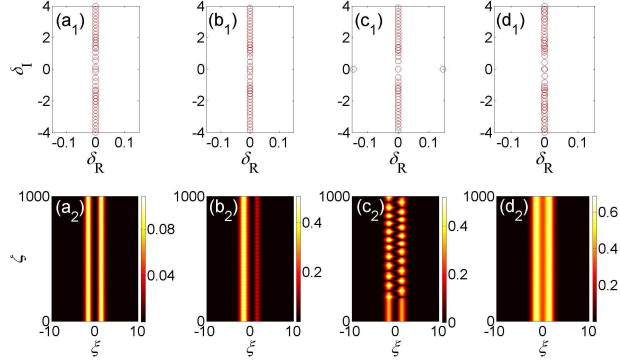} \vspace{%
0cm}
\caption{(Color online) The top row: the same as in Fig. \protect\ref%
{figure10}, but for the self-defocusing nonlinearity, $\protect\sigma =-1$.
The bottom row: perturbed evolution of the solitons from the top row. (a$_{1}
$,a$_{2}$): A stable antisymmetric soliton with $\protect\beta =1.3$. Panels
(b$_{1}$,b$_{2}$), (c$_{1}$,c$_{2}$), and (d$_{1}$,d$_{2}$) display stable
asymmetric, unstable antisymmetric and stable symmetric soliton solutions,
respectively, at $\protect\beta =1.12$.}
\label{figure11}
\end{figure}

\begin{figure}[tbp]
\centering\vspace{0cm} \includegraphics[width=8.5cm]{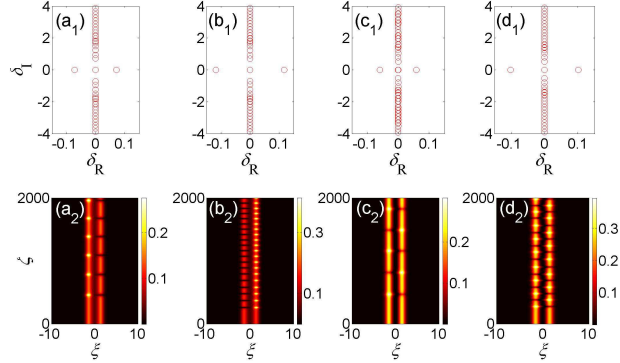} \vspace{%
0cm}
\caption{(Color online) The top row: linear-stability spectra for weakly
unstable soliton solutions, taken close to the SSB bifurcation points. The
bottom row: the perturbed evolution of the solitons from the top row. Panels
(a$_{1}$,a$_{2}$) and (b$_{1}$,b$_{2}$) display unstable symmetric solitons
with $\protect\beta =1.65$ and $1.67$, respectively, in the case of
self-focusing. Panels (c$_{1}$,c$_{2}$) and (d$_{1}$,d$_{2}$) display
unstable antisymmetric solitons with $\protect\beta =1.2$ and $1.17$,
respectively, in the case of self-defocusing.}
\label{figure12}
\end{figure}

To corroborate predictions of the linear-stability analysis, we have
performed numerical simulations of perturbed evolution of the solitons. We
start with the self-focusing nonlinearity and display the result for a
stable symmetric soliton at $\beta =1.62<\beta _{\mathrm{cr}}=1.6389$. The
respective eigenvalue spectrum is shown in Fig. \ref{figure10}(a$_{1}$), and
the nonlinear evolution under the action of random-noise initial
perturbations with a relative amplitude of $10\%$ is presented in Fig. \ref%
{figure10}(a$_{2}$). It is seen that, at least up to $\zeta =1000$, the
solution indeed remains stable. Next, we address stable asymmetric, unstable
symmetric, and stable antisymmetric soliton solutions for $\beta =1.75>\beta
_{\mathrm{cr}}=1.6389$ [stationary profiles of these solutions are shown in
Figs. \ref{figure1}(b), \ref{figure1}(c), and \ref{figure1}(d),
respectively]. The corresponding linear-stability spectra are shown in top
panels of Figs. \ref{figure10}(b$_{1}$), \ref{figure10}(c$_{1}$), and \ref%
{figure10}(d$_{1}$). The evolution of the stable asymmetric and stable
antisymmetric solitons, under the action of random perturbations with the $%
10\%$ relative amplitude, is displayed in Figs. \ref{figure10}(b$_{2}$) and %
\ref{figure10}(d$_{2}$), respectively. On the other hand, Fig. \ref{figure10}%
(c$_{2}$) demonstrates that the symmetric soliton, whose instability is
predicted by the spectrum shown in Fig. \ref{figure10}(c$_{1}$), quickly
develops the instability (even without adding a perturbation to the initial
state), and, as it might be expected, the instability tends to convert it
into a stable asymmetric state.

In Fig. \ref{figure11}, we display the evolution of the soliton solutions in
the case of the self-defocusing nonlinearity, where an antisymmetric soliton
at $\beta =1.3>\beta _{\mathrm{cr}}=1.2138$, which is predicted to be stable
by Fig. \ref{figure11}(a$_{1}$) [$\beta >\beta _{\mathrm{cr}}$ is the
stability condition in the case of the self-defocusing nonlinearity,
according to Fig. \ref{figure2}(a)], indeed keeps its shape in the course of
the perturbed evolution in Fig. \ref{figure11}(a$_{2}$). For stable
asymmetric, unstable antisymmetric, and stable symmetric solitons, with $%
\beta =1.12<\beta _{\mathrm{cr}}=1.2138$ [stationary profiles of these
solutions are shown in Figs. \ref{figure2}(b), \ref{figure2}(c), and \ref%
{figure2}(d), respectively], the linear-stability spectra are shown in Figs. %
\ref{figure11}(b$_{1}$), \ref{figure11}(c$_{1}$), and \ref{figure11}(d$_{1}$%
), respectively. The perturbed evolution of the solutions, displayed in
Figs. \ref{figure11}(b$_{2}$) and \ref{figure11}(d$_{2}$), corroborates the
predicted stability. On the other hand, the unstable antisymmetric soliton
develops the instability in the numerical simulations even without addition
of perturbations, as seen in Fig. \ref{figure11}(c$_{2}$). It is worthy to
note that, in this case, the evolution does not tend to convert the unstable
soliton into an asymmetric one that would be spontaneously pinned to one
potential well; instead, it develops periodic oscillations between the two
wells.

We have also considered the evolution of unstable soliton solutions near the
bifurcation points, in the cases of both the self-focusing and defocusing
signs of the nonlinearity. The results are summarized in Fig. \ref{figure12}%
. Without the addition of initial perturbations, these unstable solitons
develop oscillations, with a period that is larger for a weaker instability
[smaller $\max \left( \delta _{R}\right) $]. Thus, the spontaneously
emerging oscillations are slower in Figs. \ref{figure12}(a$_{2}$) and \ref%
{figure12}(c$_{2}$) than in Figs. \ref{figure12}(b$_{2}$) and \ref{figure12}%
(d$_{2}$).

\section{Conclusion}

We have investigated the phenomenology of SSB (spontaneous symmetry
breaking) of spatial Kerr solitons in the model based on the NLFSE
(nonlinear fractional Schr\"{o}dinger equation) with a symmetric double-well
potential and self-focusing or defocusing cubic nonlinearity. In agreement
with the general principles of the SSB theory, the increase of the soliton's
integral power (norm) destabilizes symmetric and antisymmetric solitons, and
creates stable asymmetric ones, in the cases of the self-focusing and
self-defocusing nonlinearities, respectively, In either case, stable
asymmetric solitons emerge at the SSB bifurcation point, the pitchfork
bifurcation always being of the supercritical (forward) type. In the case of
the self-defocusing nonlinearity, the bifurcation may be considered as an
inverted one, in the sense that it happens with the decrease of the
soliton's propagation constant. The dependence of the SSB effects on the L%
\'{e}vy index, $\alpha $, which characterizes the fractional dimension in
the NLFSE, and separation, $\xi _{0}$, between the potential wells in the
symmetric potential, has been explored. In particular, the increase of $%
\alpha $ leads to shrinkage or expansion of the SSB area, depending on the
value of $\xi _{0}$.

A relevant direction for the extension of the analysis is to develop it for
the study of SSB phenomena in $\mathcal{PT}$-symmetric and dissipative
systems.

\section{Acknowledgments}

This work was supported by the National Natural Science Foundation of China (NNSFC) (11805141, 11705108, 11804246), Shanxi Province Science Foundation for Youths (201901D211424), Scientific and Technological Innovation Programs of Higher Education Institutions in Shanxi (STIP) (2019L0782), and Zhejiang Provincial Natural Science Foundation of China (Grant no. LR20A050001). The work of BAM
was supported, in a part, by the Israel Science Foundation through grant No.
1287/17.

\end{document}